Fig. 1

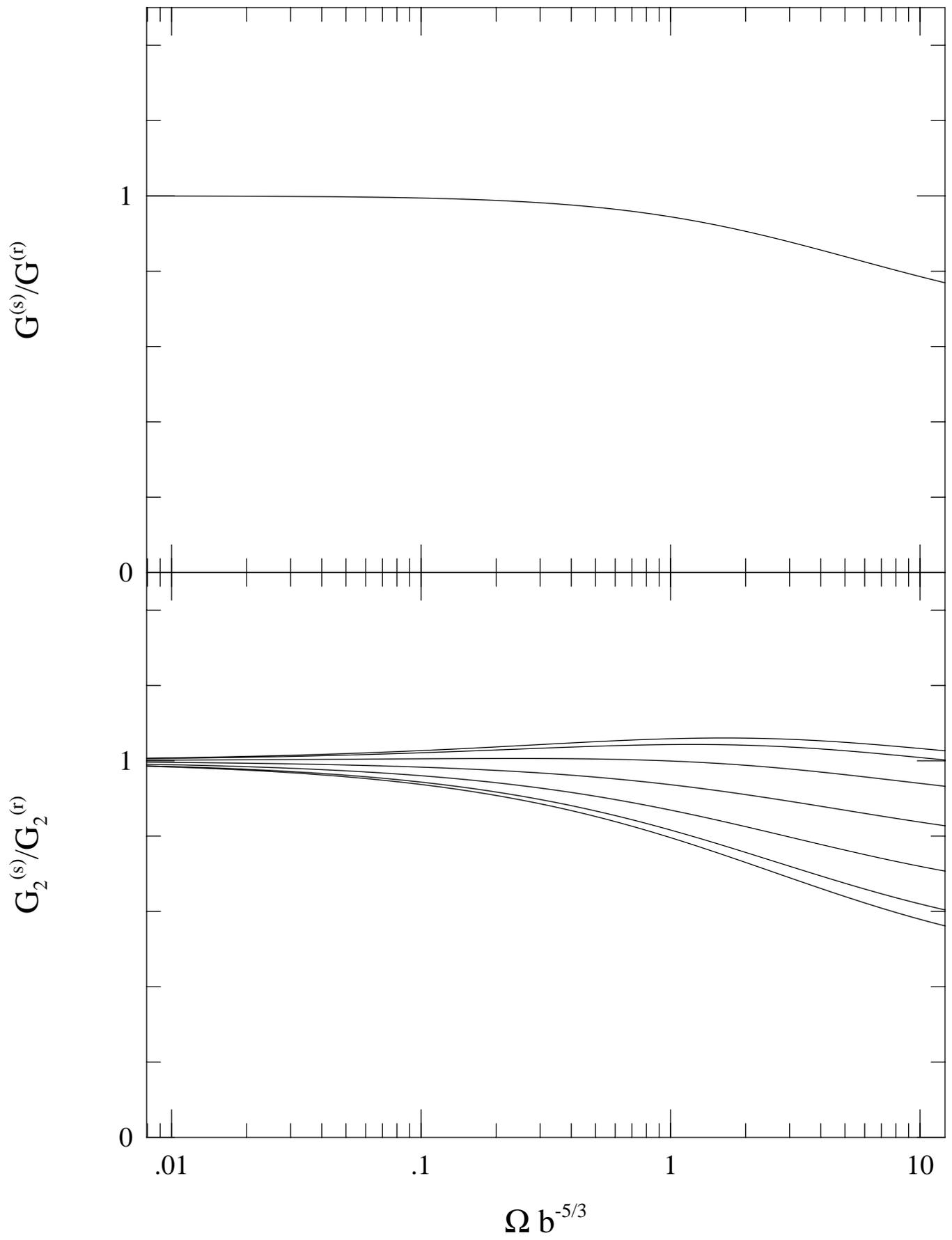

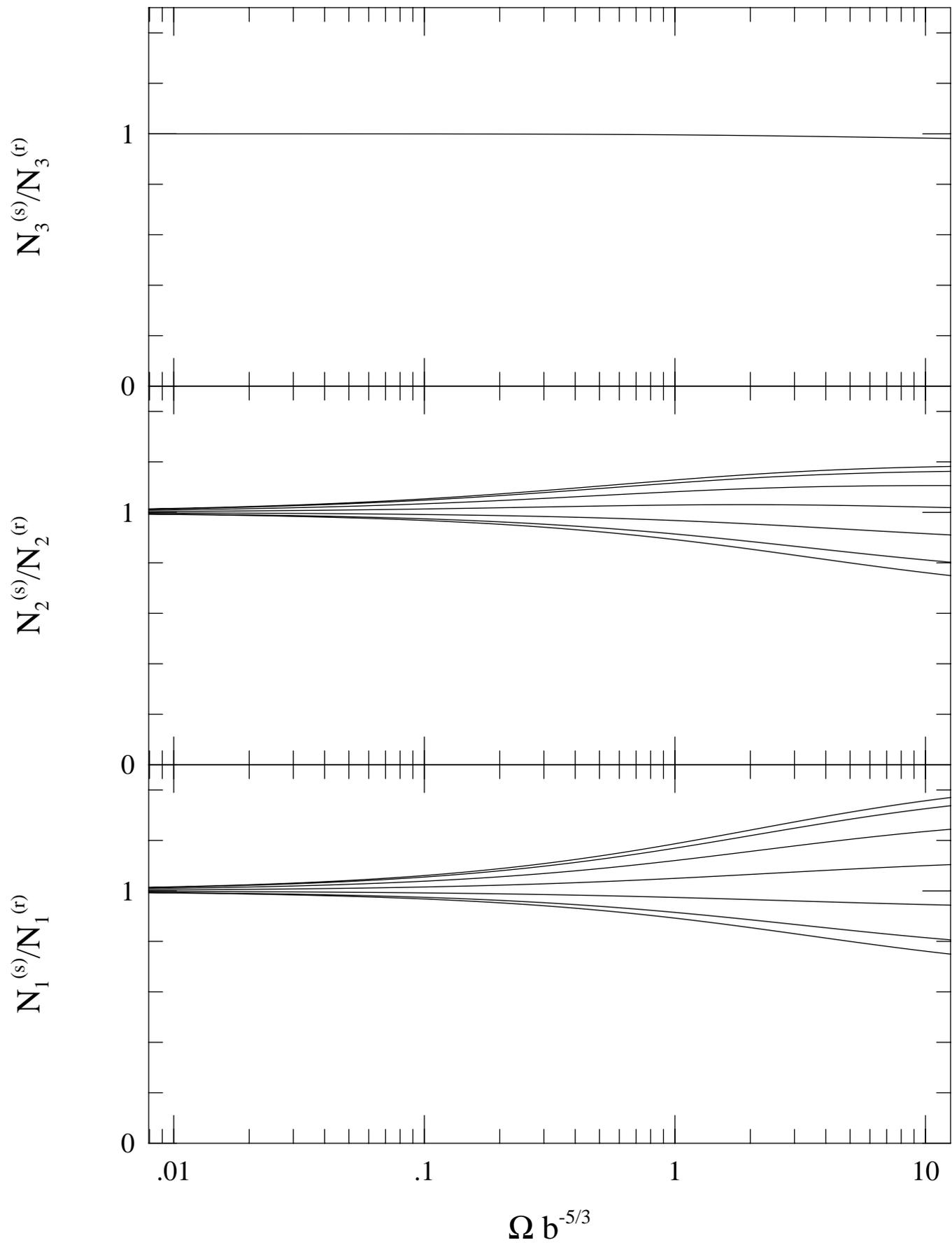

Fig. 2

Fig. 3

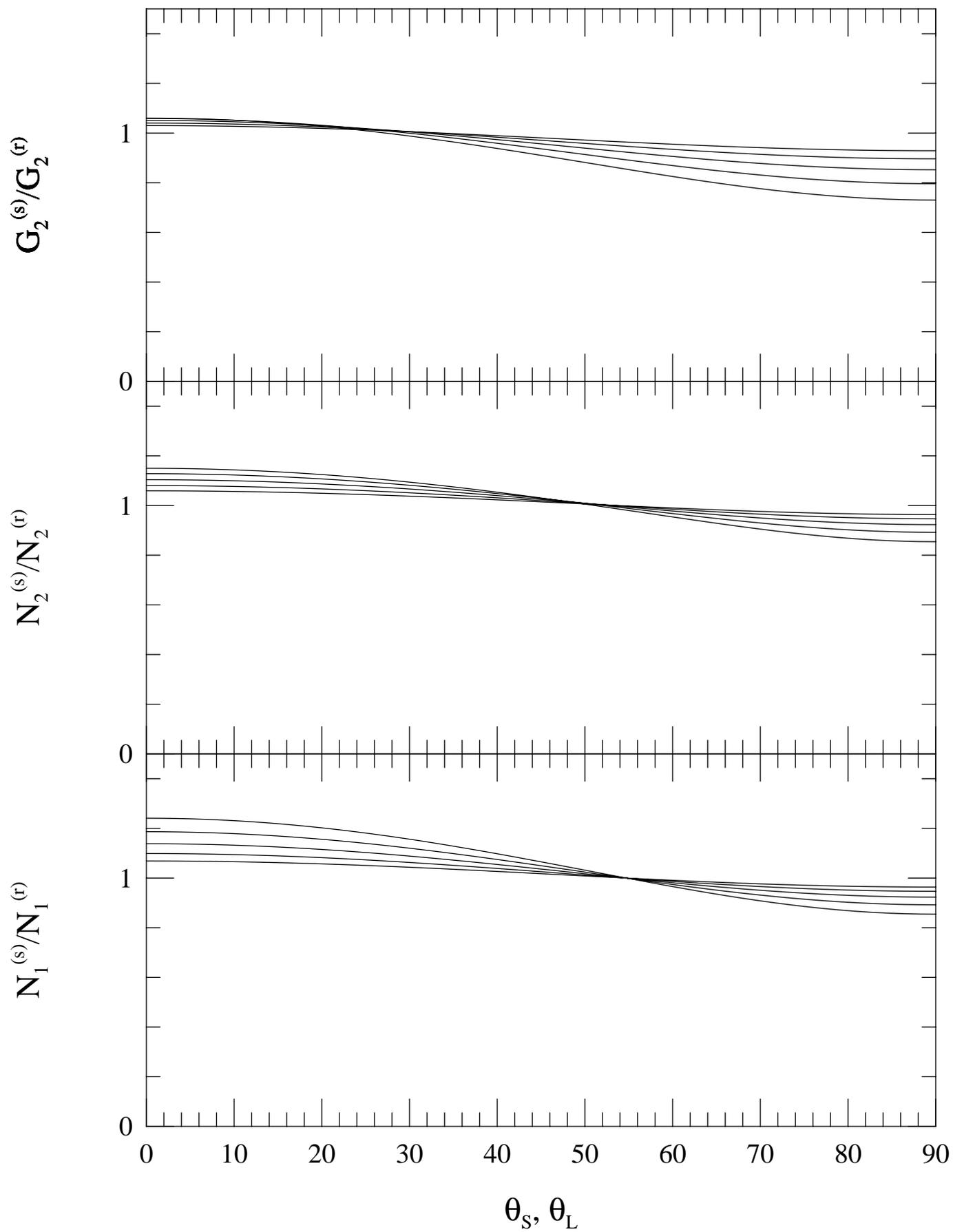

$\theta_S, \theta_L$

# STATISTICS OF ISODENSITY CONTOURS IN REDSHIFT SPACE

Takahiko Matsubara[*]

Department of Physics
The University of Tokyo
Bunkyo-ku, Tokyo 113, Japan

and

Department of Physics
Hiroshima University
Higashi-Hiroshima 724, Japan.

## Abstract

The peculiar velocities of galaxies distort the clustering pattern anisotropically in redshift space. This effect on the statistics of isodensity contours is examined by linear theory. The statistics considered in this paper are the three- and two-dimensional genus of isodensity contours, the area of isodensity contours, the length of isodensity contours in the 2-dimensional slice and the level crossing statistics on the line. We find that all these statistics in redshift space as functions of density threshold of contours have the same shape as in real space. The redshift space distortion affects only amplitudes of these statistics. The three-dimensional genus and the area hardly suffer from the redshift space distortion for $0 \leq \Omega b^{-5/3} \leq 1$, where $b$ is a linear bias parameter. The other statistics are defined in one- or two-dimensional slices of the sample volume and depend on the direction of these slices relative to the line of sight. This dependence on direction of these statistics provides a way to determine the density parameter of the universe.

*Subject headings*: cosmology: theory — galaxies: clustering — galaxies: distances and redshifts — gravitation

[*]E-mail address: matsu@yayoi.phys.s.u-tokyo.ac.jp

# 1. INTRODUCTION

Redshift surveys of galaxies play essential roles in revealing the structure of our universe. If galaxies move purely with the uniform Hubble expansion, redshift surveys would tell us the real distribution of galaxies. In reality, however, the peculiar velocities of galaxies distort the distribution in mapping from real space to redshift space. The distortion is along the line of sight and the clustering pattern of galaxies in redshift space becomes anisotropic.

There are two characteristic features in the redshift space distortion. On very small scales, the random peculiar motions in virializing clusters stretch the shape of clusters along the line of sight: the 'finger of God' effect. As a result, the strength of the clustering is weaker in redshift space than in real space (e.g., Lilje & Efstathiou 1989; Suto & Suginohara 1991; Peacock 1993; Matsubara 1994a). On large scales, the coherent velocity field falling in the region with the excess mass make the perturbation enhanced along the line of sight, in contrast to the small-scale case (Sargent & Turner 1977; Kaiser 1987; Lilje & Efstathiou 1989; McGill 1990).

The redshift data are to be compared with many theories on the structure formation of the universe. The straightforward predictions of these theories, however, are usually described in real space. To compare the theories with observations, the various statistical measures, such as correlation functions, probability distribution function etc. are used. Among such statistics, there is a class of statistics using a smoothed density field which cut the noisy property of galaxy distribution.

For example, Gott, Melott & Dickinson (1986) introduced the topology of isodensity contours of those smoothed field. The genus $G$, which is defined by $-1/2$ times the Euler characteristics of two-dimensional surfaces, can be a quantitative measure of the topology and was analyzed both in numerical simulations and in redshift surveys of galaxies by many authors (Gott, Weinberg & Melott 1987; Weinberg, Gott & Melott 1987; Melott, Weinberg & Gott 1988; Gott et al. 1989; Park & Gott 1991; Park, Gott & da Costa 1992; Weinberg & Cole 1992; Moore et al. 1992; Vogeley, Park, Geller, Huchra & Gott 1994; Rhoads, Gott & Postman 1994). Analytic expressions of the genus for some cases are known so far, including Gaussian random field (Doroshkevich 1970; Adler 1981; Bardeen et al. 1986; Hamilton, Gott & Weinberg 1986), Rayleigh-Lévy random-walk fractal (Hamilton 1988), union of overlapping balls (Okun 1990) and weakly non-Gaussian random fields (Matsubara 1994b). So far these expressions have been derived only for isotropic fields.

There are other statistics of isodensity contours, which include the 2D genus in two-dimensional slices of density field $G_2$ (Melott et al. 1989), the area of isodensity contours $N_3$, the length of isodensity contours in two-dimensional slices of density field $N_2$ and the level crossing statistics $N_1$ (Ryden 1988; Ryden et al. 1989). The analytic expressions of all the above statistics for isotropic Gaussian random fields are known.

The growth of the density fluctuation of the universe on large scales is described by linear theory in the gravitational instability picture of the structure formation (e.g., Peebles 1980). Thus, if the initial fluctuation is a Gaussian random field as is often assumed, the statistics of isodensity contours of the density field with large smoothing length should obey the random Gaussian prediction. Because the known analytic expression for Gaussian random fields is for the isotropic field, this Gaussianity test of the initial fluctuation should be performed in real space which is not feasible in reality. It is not obvious whether or not the redshift space distortion strongly affects statistics of isodensity contours. As for the genus, Melott, Weinberg & Gott (1988) found by analysis of $N$-body simulations that genus is hardly affected by redshift space distortion when the smoothing length is larger than the correlation length $\sim 5h^{-1}$Mpc.

In this paper, the redshift space distortions of statistics of isodensity contours $G$, $G_2$, $N_3$, $N_2$, $N_1$ are studied analytically by linear theory of gravitational instability assuming that the initial fluctuation is a Gaussian random field. These statistics will be accurately determined with the future redshift surveys. Our approach provides the Gaussianity test which can be directly performed in redshift space. Moreover, the redshift space distortion generally depends on the density parameter of the universe and our formula could in principle discriminate this parameter.

The rest of this paper is organized as follows. In §2, the distant-observer approximation to evaluate the redshift space distortion of statistics is introduced. We show directly that our definition of this approximation is equivalent to the approximation adopted by Kaiser (1987) who gave for the first time the redshift space distortion of two-point statistics by linear theory. Then we derive the useful anisotropic statistics in redshift space. The main results of this paper, the formula for statistics of isodensity contours in redshift space, are derived in §3. We discuss the results in §4.

## 2. FIELD CORRELATIONS IN THE DISTANT-OBSERVER APPROXIMATION

Kaiser (1987) showed that the distortion of power spectrum in redshift space $P^{(s)}(\boldsymbol{k})$ from that in real space $P^{(r)}(k)$ is given by the simple formula as

$$P^{(s)}(\boldsymbol{k}) = \left[1 + f\mu^2\right]^2 P^{(r)}(k), \tag{2.1}$$

where $\mu$ is the cosine of the angle between the line of sight and the direction of $\boldsymbol{k}$ and $f(\Omega) = H^{-1}\dot{D}/D \approx \Omega^{0.6}$, $D$ is the linear growth rate and $H$ is the Hubble parameter. The omega-dependence of $f(\Omega)$ is approximately the same in the presence of cosmological constant $\Lambda$ (see, for detail, Lahav et al. 1991), and we use this approximation, $f \approx \Omega^{3/5}$, extensively in this paper. This simplicity of equation (2.1) relies on the approximation that the sample volume is distant from the observer. Inhomogeneity of the redshift samples closer to the observer is not negligible as well as the anisotropy and prevent to give the simple expression as in equation (2.1). When the sample volume is distant from the observer, the direction of line of sight is approximately fixed in the sample volume. We call this approximation fixing the line of sight as 'distant-observer approximation'. Adopting this approximation, the Cartesian coordinates in which the line of sight is fixed is convenient for our purpose. The Cartesian coordinates make the calculation of statistics of isodensity contours easy. Our distant-observer approximation exactly reproduces Kaiser's result (eq.[2.1]) which is derived by first introducing the spherical coordinates and then approximating that the sample is distant from the observer. Our approach depends on the Cartesian coordinates from the beginning and it would be useful to see directly the equivalence of the Kaiser's approximation and our distant-observer approximation. The derivation of Kaiser's result in Cartesian coordinates is simpler as we will see in the following.

In the Cartesian coordinates of our distant-observer approximation, we define the direction of the line of sight by an unit vector $\hat{\boldsymbol{z}}$. Using the line-of-sight component of a peculiar velocity field $U(\boldsymbol{r}) = \boldsymbol{v}(\boldsymbol{r}) \cdot \hat{\boldsymbol{z}}$, the mapping of the coordinates from the real space to the redshift space is given by

$$\boldsymbol{s}(\boldsymbol{r}) = \boldsymbol{r} + \frac{\hat{\boldsymbol{z}}}{H}[U(\boldsymbol{r}) - U(\boldsymbol{0})]. \tag{2.2}$$

The observer is placed on the origin of the coordinates, $\boldsymbol{0}$. On large scales we are interested in, we can relate the number density of galaxies in redshift space $\rho_g^{(s)}$ and that in real space $\rho_g^{(r)}$ by evaluating the Jacobian of the mapping (2.2) resulting in

$$\rho_g^{(s)}(\boldsymbol{s}(\boldsymbol{r})) = \left(1 + \frac{1}{H}\hat{\boldsymbol{z}} \cdot \nabla U(\boldsymbol{r})\right)^{-1} \rho_g^{(r)}(\boldsymbol{r}). \tag{2.3}$$

Leaving only linear order in density contrast $\delta = \rho/\bar{\rho} - 1$ and peculiar velocity field, this relation reduces to

$$\delta_g^{(s)}(\boldsymbol{r}) = \delta_g^{(r)}(\boldsymbol{r}) - \frac{1}{H}\hat{\boldsymbol{z}} \cdot \nabla U(\boldsymbol{r}). \tag{2.4}$$

The peculiar velocity field in linear theory (Peebles 1980) is, in growing mode,

$$\boldsymbol{v}(\boldsymbol{r}) = -Hf\nabla\triangle^{-1}\delta_m^{(r)}(\boldsymbol{r}), \tag{2.5}$$

where $\triangle^{-1}$ is the inverse Laplacian and $\delta_m^{(r)}$ is the mass density contrast in real space. In the following, $\delta_g$ and $\delta_m$ are assumed to be proportional to each other. This assumption is called linear biasing: $\delta_g = b\delta_m$, where $b$ is the bias parameter which is a constant. The relation between the density contrast in redshift space and in real space is, up to linear order,

$$\delta_g^{(s)}(\boldsymbol{r}) = \left[1 + f\, b^{-1}(\hat{\boldsymbol{z}} \cdot \nabla)^2 \triangle^{-1}\right] \delta_g^{(r)}(\boldsymbol{r}), \tag{2.6}$$

or, in Fourier space,

$$\widetilde{\delta}_g^{(s)}(\boldsymbol{k}) = \left[1 + f\, b^{-1}\left(\frac{\hat{\boldsymbol{z}} \cdot \boldsymbol{k}}{k}\right)^2\right] \widetilde{\delta}_g^{(r)}(\boldsymbol{k}), \tag{2.7}$$

which is Kaiser's result.

In the following, we use the parameters $\sigma_j$ and $C_j$ defined by

$$\sigma_j^2(R) = \int \frac{k^2 dk}{2\pi^2} k^{2j} P^{(r)}(k) W^2(kR), \tag{2.8}$$

$$C_j(\Omega) = \frac{1}{2}\int_{-1}^{1} d\mu\, \mu^{2j}\left(1 + f\, b^{-1}\mu^2\right)^2, \tag{2.9}$$

where $W(x)$ is the Fourier transform of the window function to smooth the noisy field of galaxy distribution. The two popular windows are the Gaussian window $W_\mathrm{G}(x) = \exp(-x^2/2)$ and the top-hat window $W_\mathrm{TH}(x) = 3(\sin x - x\cos x)/x^3$. We assume that the window function is an isotropic function. The $rms$ $\sigma^{(s)}$ of density contrast in redshift space is given by

$$\left(\sigma^{(s)}\right)^2 = C_0 \sigma_0^2. \tag{2.10}$$

We define the following normalized quantities,

$$\alpha = \frac{\delta_R^{(s)}}{\sigma^{(s)}}, \quad \beta_i = \frac{\partial_i \delta_R^{(s)}}{\sigma^{(s)}}, \quad \omega_{ij} = \frac{\partial_i \partial_j \delta_R^{(s)}}{\sigma^{(s)}}, \tag{2.11}$$

– 5 –

where $\delta_R^{(s)}$ is the smoothed density contrast in redshift space. These quantities obey the multivariate Gaussian distribution in linear theory if the primordial fluctuation is a random Gaussian field. The multivariate Gaussian distribution is completely determined by the correlations of all pairs of variables. For our purpose below, the statistics of quantities $\alpha$, $\beta_i$, $\omega_{IJ}$ ($i = 1, 2, 3$; $I, J = 1, 2$) is sufficient. Choosing the coordinates in which the line of sight is the third axis, all the correlations among the above quantities at some point are as follows:

$$\langle \alpha \alpha \rangle = 1,$$
$$\langle \alpha \beta_i \rangle = 0,$$
$$\langle \alpha \omega_{IJ} \rangle = \frac{1}{2} \left( \frac{C_1}{C_0} - 1 \right) \frac{\sigma_1^2}{\sigma_0^2} \delta_{IJ},$$
$$\langle \beta_I \beta_J \rangle = \frac{1}{2} \left( 1 - \frac{C_1}{C_0} \right) \frac{\sigma_1^2}{\sigma_0^2} \delta_{IJ},$$
$$\langle \beta_I \beta_3 \rangle = 0,$$
$$\langle \beta_3 \beta_3 \rangle = \frac{C_1}{C_0} \frac{\sigma_1^2}{\sigma_0^2},$$
$$\langle \beta_i \omega_{IJ} \rangle = 0,$$
$$\langle \omega_{IJ} \omega_{KL} \rangle = \frac{1}{8} \left( 1 - \frac{2C_1}{C_0} + \frac{C_2}{C_0} \right) \frac{\sigma_2^2}{\sigma_0^2} \left( \delta_{IJ}\delta_{KL} + \delta_{IK}\delta_{JL} + \delta_{IL}\delta_{JK} \right).$$

It is more convenient to consider

$$\tilde{\omega}_{IJ} = \omega_{IJ} - \alpha \langle \alpha \omega_{IJ} \rangle, \tag{2.12}$$

instead of $\omega_{IJ}$. The new set of variables $\alpha$, $\beta_i$, $\tilde{\omega}_{IJ}$ distribute as multivariate Gaussian and the non-vanishing correlations of these variables are only

$$\langle \alpha \alpha \rangle = 1,$$
$$\langle \beta_1 \beta_1 \rangle = \langle \beta_2 \beta_2 \rangle = \frac{1}{2} \left( 1 - \frac{C_1}{C_0} \right) \frac{\sigma_1^2}{\sigma_0^2},$$
$$\langle \beta_3 \beta_3 \rangle = \frac{C_1}{C_0} \frac{\sigma_1^2}{\sigma_0^2},$$
$$\langle \tilde{\omega}_{11} \tilde{\omega}_{11} \rangle = \langle \tilde{\omega}_{22} \tilde{\omega}_{22} \rangle = \frac{1}{8} \left[ 3 \left( 1 - \frac{2C_1}{C_0} + \frac{C_2}{C_0} \right) - 2 \left( 1 - \frac{C_1}{C_0} \right)^2 \gamma^2 \right] \frac{\sigma_2^2}{\sigma_0^2},$$
$$\langle \tilde{\omega}_{11} \tilde{\omega}_{22} \rangle = \frac{1}{8} \left[ \left( 1 - \frac{2C_1}{C_0} + \frac{C_2}{C_0} \right) - 2 \left( 1 - \frac{C_1}{C_0} \right)^2 \gamma^2 \right] \frac{\sigma_2^2}{\sigma_0^2},$$
$$\langle \tilde{\omega}_{12} \tilde{\omega}_{12} \rangle = \frac{1}{8} \left( 1 - \frac{2C_1}{C_0} + \frac{C_2}{C_0} \right) \frac{\sigma_2^2}{\sigma_0^2},$$

where $\gamma = \sigma_1^2/(\sigma_0\sigma_2)$.

## 3. STATISTICS OF ISODENSITY CONTOURS

Let us derive the formula of statistics of isodensity contours in redshift space. In the following, the primordial fluctuation is assumed to be a random Gaussian field.

### 3.1. Genus Statistics

The genus is minus a half times the Euler number. The Euler number density of isodensity contours is evaluated by

$$\text{number of maxima} + \text{number of minima} - \text{number of saddle points} \tag{3.1}$$

of the contour surfaces with regard to some fixed direction. The expectation value of Euler number of the isodensity contours per unit volume is (Doroshkevich 1970; Adler 1981; Bardeen et al. 1986)

$$n_\chi^{(s)}(\nu) = \left\langle \delta(\alpha - \nu)\delta(\beta_1)\delta(\beta_2)|\beta_3|(\omega_{11}\omega_{22} - \omega_{12}^2) \right\rangle, \tag{3.2}$$

where the isodensity contours are defined to be the surface $\delta_R^{(s)} = \nu\sigma^{(s)}$. This expression is valid even for general anisotropic fields. Using new variables $\widetilde{\omega}_{IJ}$ in equation (3.2), the following result for the expression of genus $G^{(s)}$ in redshift space is derived:

$$G^{(s)}(\nu) = -\frac{1}{2}n_\chi^{(s)}(\nu) = \frac{3\sqrt{3}}{2}\sqrt{\frac{C_1}{C_0}}\left(1 - \frac{C_1}{C_0}\right)G^{(r)}(\nu), \tag{3.3}$$

where, from equation (2.9),

$$\frac{C_1}{C_0} = \frac{1}{3}\frac{1 + \frac{6}{5}fb^{-1} + \frac{3}{7}(fb^{-1})^2}{1 + \frac{2}{3}fb^{-1} + \frac{1}{5}(fb^{-1})^2}, \tag{3.4}$$

and $G^{(r)}$ is genus in real space (Doroshkevich 1970; Adler 1980; Bardeen et al. 1986; Hamilton et al. 1986) given by

$$G^{(r)}(\nu) = \frac{1}{(2\pi)^2}\left(\frac{\sigma_1}{\sqrt{3}\sigma_0}\right)^3(1 - \nu^2)e^{-\nu^2/2}. \tag{3.5}$$

The redshift space distortion does not alter the shape of genus as a function of density threshold and only the amplitude is affected. The $\Omega$ dependence of the change in amplitude is plotted in Fig. 1 (upper panel). The effects of redshift space distortion is small for $\Omega b^{-5/3}$ less than unity. This fact is in agreement with the $N$-body analysis of Melott, Weinberg & Gott (1988).

### 3.2. 2D Genus Statistics

The next statistics we consider is 2-dimensional genus. This statistics is defined in the 2-dimensional flat plane $S$ in 3-dimensional space. The density field calculated in 3-dimensional volume defines the high density points in the plane which constitute the excursion set on the plane. The 2D genus is defined by the number of contours surrounding high density region minus the number of contours surrounding low density regions (Adler 1980; Coles 1988; Melott et al 1989; Gott et al. 1990). The redshift space is anisotropic by the presence of the special direction, line of sight, so the 2D statistics depends on the angle $\theta_S$ between the plane $S$ and the line of sight. The alternative, equivalent definition of 2D genus is useful in the following. For some arbitrarily fixed direction in the 2D surface, the maximum and minimum points are defined on the contours. These points are classified into upcrossing and downcrossing points with respect to the fixed direction. The 2D genus is defined to be

$$\frac{1}{2}(\text{number of upcrossing minima} - \text{number of upcrossing maxima}$$
$$- \text{number of downcrossing minima} + \text{number of downcrossing maxima}),$$
(3.6)

of the contour lines with regard to some fixed direction in the plane $S$. The latter definition can be used to obtain the following expression for 2D genus per unit area of the plane:

$$G_2^{(s)}(\nu, \theta_S) = -\frac{1}{2} \left\langle \delta(\alpha - \nu)\delta(\beta_1)|\beta_2 \sin\theta_S + \beta_3 \cos\theta_S|\omega_{11} \right\rangle. \tag{3.7}$$

The corresponding expression in case of the isotropic 2-dimensional field is appeared in Bond & Efstathiou (1987). From this expression, we obtain

$$G_2^{(s)}(\nu, \theta_S) = \frac{3}{2}\sqrt{\left(1 - \frac{C_1}{C_0}\right)\left[1 - \frac{C_1}{C_0} + \left(\frac{3C_1}{C_0} - 1\right)\cos^2\theta_S\right]} G_2^{(r)}(\nu). \tag{3.8}$$

To derive this result, we use $\tilde{\omega}_{11}$ rather than $\omega_{11}$, then regard the variables $\alpha$, $\beta_1$, $\beta_2 \sin\theta_S + \beta_3 \cos\theta_S$ and $\tilde{\omega}_{11}$ as independent variables. The 2D genus in real space $G_2^{(r)}$ has the following form:

$$G_2^{(r)}(\nu) = \frac{1}{(2\pi)^{3/2}} \left(\frac{\sigma_1}{\sqrt{3}\sigma_0}\right)^2 \nu e^{-\nu^2/2}. \tag{3.9}$$

The redshift space distortion again affects only amplitude of 2D genus. The dependence on $\Omega$ and $\theta_S$ of the change in amplitude is plotted in Fig. 1 (lower panel). The dependence on the direction of the plane $\theta_S$ for large $\Omega$ can be used to determine the cosmological parameter by this statistics.

### 3.3. Area of Isodensity Contours

The area of isodensity contours per unit volume (Ryden 1988; Ryden et al. 1989) is given by

$$N_3^{(s)}(\nu) = \left\langle \delta(\alpha - \nu)\sqrt{\beta_1^2 + \beta_2^2 + \beta_3^2} \right\rangle. \tag{3.10}$$

This expression is valid even for general anisotropic fields. Introducing spherical coordinates for $\beta_i$, the above expression is calculated to be

$$N_3^{(s)}(\nu) = \frac{\sqrt{3}}{2} \left[ \sqrt{\frac{C_1}{C_0}} - \frac{1 - \dfrac{C_1}{C_0}}{2\sqrt{2\left(\dfrac{3C_1}{C_0} - 1\right)}} \ln \left| \frac{\sqrt{\dfrac{C_1}{C_0}} - \sqrt{\dfrac{1}{2}\left(\dfrac{3C_1}{C_0} - 1\right)}}{\sqrt{\dfrac{C_1}{C_0}} + \sqrt{\dfrac{1}{2}\left(\dfrac{3C_1}{C_0} - 1\right)}} \right| \right] N_3^{(r)}(\nu), \tag{3.11}$$

where $N_3^{(r)}$ is the area in real space given by

$$N_3^{(r)}(\nu) = \frac{2}{\sqrt{3}\pi} \frac{\sigma_1}{\sigma_0} e^{-\nu^2/2}. \tag{3.12}$$

Again, only the amplitude is affected. The $\Omega$ dependence of the amplitude is very weak as plotted in Fig. 2 (upper panel).

### 3.4. Length of Isodensity Contours in Planes

As in the case of 2D genus statistics, 2-dimensional flat plane $S$ is considered in the statistics of length of isodensity contours. The length of intersections of isodensity contours and the plane $S$ was introduced by Ryden (1988). For isotropic density fields, this statistics is proportional to the area statistics considered in the previous section. As shown below, for anisotropic fields in redshift space, the proportional factor depends on the direction of the surface $S$ relative to the line of sight. The angle $\theta_S$ between the plane $S$ and the line of sight is relevant as in the case of 2D genus statistics. The expectation value of length of isodensity contours in the plane $S$ per unit area of the plane is given by

$$N_2^{(s)}(\nu, \theta_S) = \left\langle \delta(\alpha - \nu)\sqrt{\beta_1^2 + (\beta_2 \sin\theta_S + \beta_3 \cos\theta_S)^2} \right\rangle. \tag{3.13}$$

To evaluate the above equation, note that $\alpha$, $\beta_1$ and $\beta_2 \sin\theta_S + \beta_3 \cos\theta_S$ are non-correlated, independent variables. Introducing the polar coordinates for the latter two variables, the following expression is obtained:

$$N_2^{(s)}(\nu, \theta_S) = \frac{\sqrt{6}}{\pi} \sqrt{1 - \frac{C_1}{C_0} + \left(\frac{3C_1}{C_0} - 1\right)\cos^2\theta_S}\, E\left( \frac{\left(\dfrac{3C_1}{C_0} - 1\right)\cos^2\theta_S}{1 - \dfrac{C_1}{C_0} + \left(\dfrac{3C_1}{C_0} - 1\right)\cos^2\theta_S} \right) N_2^{(r)}(\nu), \tag{3.14}$$

where $E(k)$ is the complete elliptical integral of the second kind:

$$E(k) = \int_0^{\pi/2} \sqrt{1 - k^2 \sin^2 \phi} d\phi, \tag{3.15}$$

and $N_2^{(r)}$ is the expectation in real space:

$$N_2^{(r)}(\nu) = \frac{\pi}{4} N_3^{(r)}. \tag{3.16}$$

The redshift space distortion affects only amplitude as other statistics considered in this paper. The dependence on $\Omega$ and $\theta_S$ of the amplitude is plotted in Fig. 2 (middle panel).

### 3.5. Contour Crossings

Contour crossing statistics is the mean number of intersection of a straight line $L$ and the isodensity contours. This statistics of large-scale structure is introduced by Ryden (1988) and extensively studied by Ryden et al. (1989) using numerical simulations and redshift observations. For isotropic density fields, this statistics is also proportional to the area statistics. In redshift space, the density field is anisotropic and this statistics depends on the angle $\theta_L$ between the direction of the line $L$ and the line of sight. The mean number of crossings per unit length of the line $L$ is given by

$$N_1^{(s)}(\nu, \theta_L) = \langle \delta(\alpha - \nu) |\beta_1 \sin\theta_L + \beta_3 \cos\theta_L| \rangle. \tag{3.17}$$

This expression is evaluated by noting that $\alpha$ and $\beta_1 \sin\theta_L + \beta_3 \cos\theta_L$ are non-correlated, independent variables, resulting in

$$N_1^{(s)}(\nu, \theta_L) = \sqrt{\frac{3}{2} \left[1 - \frac{C_1}{C_0} + \left(\frac{3C_1}{C_0} - 1\right) \cos^2\theta_L\right]} N_1^{(r)}(\nu), \tag{3.18}$$

where $N_1^{(r)}$ is the expectation in real space:

$$N_1^{(r)}(\nu) = \frac{1}{2} N_3^{(r)}(\nu). \tag{3.19}$$

Again, the redshift space distortion affects only amplitude. The dependence on $\Omega$ and $\theta_L$ of the amplitude is plotted in Fig. 2 (lower panel).

## 4. DISCUSSION

The strength of the effects of redshift space distortion is different according to which statistics is focused on. The characteristic point in linear theory is that all the statistics in redshift space considered in this paper has the same shapes as in real space as functions of density threshold. The redshift space distortion affects only on the amplitude.

As for genus $G(\nu)$ and area $N_3(\nu)$, the redshift distortion of amplitude is small for $0 \leq \Omega b^{-5/3} \leq 1$ (Fig. 1, 2). This property justifies the comparison of the observational redshift data and the theoretical Gaussian prediction in real space [eq. (3.5) and eq. (3.12)] at least in linear regime. In the Gaussianity test of primordial fluctuation using genus and area statistics, the effect of redshift space distortion can be ignored approximately. The similar statistical measure of galaxy clustering, the skewness $\langle \delta_R^3 \rangle / \langle \delta_R^2 \rangle^2$ induced by weakly nonlinear evolution from the Gaussian primordial fluctuation is recently reported not to be affected much also by redshift space distortion (Juszkiewicz, Bouchet, & Colombi 1993; Hivon et al. 1994).

The direction dependent statistics, 2D genus $G_2$, length statistics $N_2$ and crossing statistics $N_1$ are shown to exhibit the dependence on $\Omega$ and the direction to define statistics. In Fig. 3 plotted the direction dependence of these three statistics. The direction dependence of amplitude of these statistics is relatively large: for $\Omega b^{-5/3} = 1$, the amplitude varies more than 20 % while for $\Omega b^{-5/3} = 0$, the amplitude does not varies and is equal to the one in real space. The direction-dependence depends on $\Omega b^{-5/3}$ and the statistics $G_2$, $N_2$ and $N_1$ are three independent indicators to determine the cosmological parameters. The redshift space distortion of power spectrum (eq. [2.1]) or two-point correlation function of Kaiser's result recently used for determining the parameter $\Omega b^{-5/3}$ (Hamilton 1992; 1993; Fry & Gaztañaga 1994; Cole, Fisher & Weinberg 1994). Gramann, Cen & Gott (1994) introduced the ratio of density gradients $\langle (\partial \delta^{(s)}/\partial r_\parallel)^2 \rangle / \langle (\partial \delta^{(s)}/\partial r_\perp)^2 \rangle$, where $r_\perp$, $r_\parallel$ are spatial component of line of sight and its perpendicular component, as a discriminator of $\Omega b^{-5/3}$. This ratio is equal to $3C_1/C_0$ in linear theory. Our results can be used as complementary ways of these observations.

The purpose of this paper is to give the theoretical relations of statistics of isodensity contours in redshift space and we did not to apply those results to real data. We are planning the applications of our results to data of $N$-body simulations and redshift observations.

I am grateful to thank Y. Suto for a careful reading of the manuscript and useful comments.

I acknowledge the support of JSPS Fellowship. This research was supported in part by the Grants-in-Aid for Scientific research from Ministry of Education, Science and Culture of Japan (No. 0042).

# FIGURE CAPTIONS

**Figure 1 :** *Upper panel*: amplitude of genus in redshift space relative to that in real space. *Lower panel*: relative amplitude of 2D genus. The angles between the slice and the line of sight are $\theta_S = 0°$, $15°$, $30°$, $45°$, $60°$, $75°$, $90°$ (from upper line to lower line).

**Figure 2 :** *Upper panel*: amplitude of area statistics in redshift space relative to that in real space. *Middle panel*: relative amplitude of length statistics. The angles between the slice and the line of sight are $\theta_S = 0°$, $15°$, $30°$, $45°$, $60°$, $75°$, $90°$ (from upper line to lower line). *Lower panel*: relative amplitude of level crossing statistics. The angles between the crossing line and the line of sight are $\theta_L = 0°$, $15°$, $30°$, $45°$, $60°$, $75°$, $90°$ (from upper line to lower line).

**Figure 3 :** Relative amplitude of direction-dependent statistics as functions of the angle between the line-of-sight and the plane or the line on which the statistics are evaluated. Five cases $\Omega b^{-5/3} = 0.125$, $0.25$, $0.5$, $1.0$, $2.0$ (from upper line to lower line at $\theta = 90°$) are plotted in each panel. *Upper panel*: 2D genus. *Middle panel*: length statistics *Lower panel*: level crossing statistics.

– 15 –